\documentclass[12pt]{iopart}

\usepackage{iopams}  
\usepackage{graphicx}
\begin{document}

\title[Efficient collisional blockade loading of single atom into a tight microtrap]{Efficient collisional blockade loading of single atom into a tight microtrap}

\author{Y H Fung and M F Andersen}

\address{Dodd-Walls Centre, Department of Physics,}
\address{University of Otago, PO Box 56, Dunedin 9016, New Zealand.}
\ead{mikkel@physics.otago.ac.nz}
\vspace{10pt}
\begin{indented}
\item[]February 2015
\end{indented}

\begin{abstract}
We show that controlled inelastic collisions can improve the single atom loading efficiency in the collisional blockade regime of optical microtraps. 
A collisional loss process where only one of the colliding atoms are lost, implemented during loading, enables us to kick out one of the atoms as soon as a second atom enters the optical microtrap. When this happens faster than the pair loss, which has limited the loading efficiency of previous experiments to about $50\%$, we experimentally observe an enhancement to $80\%$. A simple analytical theory predicts the loading dynamics.
Our results opens up an efficient and fast route for loading individual atoms into optical tweezers and arrays of microtraps that are too tight for easy implementation of the method reported in \cite{Andersen0, Carpentier13}. The loading of tight traps with single atoms is a requirement for their applications in future experiments in quantum information processing and few-body physics.
\end{abstract}

%
%
%
%
%

\section{Introduction}\label{S1}

Deterministic preparation of single neutral atoms in optical traps is a subject of interest due to their potential use in quantum logic devices. For more than a decade, different avenues have been pursued to achieve this goal, exploring different trap geometries and parameters, suitable for the various purposes for having the trapped single atoms. Single atoms can be isolated using the ``Rydberg blockade" \cite{Ebert14} or from quantum degenerate gases \cite{Bloch1, Greiner, Serwane11, Manning14} and individual atoms can be sorted in an array of small traps \cite{Miro06}. 

A popular approach to trap individual neutral atoms for atomic physics experiments is by using Far Off Resonance optical Traps (FORT). A FORT can, for example, be formed by tightly focussing a single laser beam, tuned far off the atomic resonance frequency. FORTs offer conservative potentials and a high level of control of a wide range of parameters. For example, the traps can be dynamically reconfigured \cite{Kaufman14}, the trapped atoms have long coherence times \cite{Schrader04}, and both the long and short range atom-atom interactions can be controlled \cite{Serwane11, Urban09, Gaetan09, Thad}. 

Several methods with various efficiencies have employed light-assisted collisions between atoms \cite{Andersen0, Carpentier13, DePueMT, Schlosser01} to ensure that a FORT site is not occupied by more than one atom. Of these, \cite{Carpentier13} reported a single atom loading efficiency of 91$\%$ in a FORT. The FORT was formed by a single laser beam focussed to a 1.8 $\mu$m spotsize, which allows for an initial sample of about $20$ atoms on average to be loaded into the FORT from a Magneto-Optical Trap (MOT). Then, a single atom isolation stage where a combination of in-trap laser cooling and light-assisted collisions induced by blue detuned light caused the atoms to be lost one by one from the FORT until only one atom remained. Naturally, a finite probability for starting the isolation stage without any atoms present will lead to a decrease in the efficiency of the method. Therefore, for this scheme to be successful, it is crucial to have an initial sample of atoms large enough such that the probability of loading zero atoms into the FORT before the isolation stage started is effectively eliminated. If the trap volume is decreased, the method will become inefficient since loading of high density atomic samples directly from a MOT is prohibited by rapid trap loss due to light-assisted collisions induced by the MOT laser light. We denote the parameter regime where the microtrap volume is so small such that it limits the efficiency of the method described in \cite{Andersen0, Carpentier13} as the ``tight microtrap regime'' \footnote{This does not correspond to a universal trap volume since in addition to the trap volume, it will depend on the loading rates that can be achieved and the parameters of the loading light used in a given experiment, but it typically occurs for trap beam waists around or below $ 1$ $\mu$m.}.

The most dramatic manifestation of the tight microtrap regime is the collisional blockade, where the volume of the microtrap is so small such that the light-assisted collisions induced by the MOT lasers cause rapid trap loss as soon as a second atom enters the trap. This prohibits or ``blockades'' the trap from being occupied by more than one atom at any stage. Light-assisted collisions induced by MOT lasers often result in both colliding atoms being lost from the trap ($2-0$ loss). This means that as soon as a second atom enters the trap, both of the colliding atoms are lost together. Previous experiments in the collisional blockade regime have therefore found approximately equal probabilities for observing 0 ($p_0$) or 1 ($p_1$) atom in the trap \cite{Schlosser01, Schlosser2} at loading times much longer than $1/R$, with $R$ being the loading rate of the tight microtrap. 

For some applications, it is a requirement that individual atoms are loaded into tight optical microtraps. For example, small microtraps are used to interface optically trapped individual atoms with microfabricated solid state structures to create hybrid devices \cite{Thompson13, Reitz13} and tight traps are favorable for ground state cooling of single atoms \cite{Kaufman12}. When collisional blockade is unavoidable, the loading efficiency of single atoms has been limited to about 50$\%$. To enhance the loading efficiency in such tight traps, one could imagine loading single atoms into traps in the regime where the method in \cite{Carpentier13} can be implemented, and then transfer the atom to a tighter trap, or in some cases it may be convenient to do a compression stage between the MOT and isolation stage as done in for example \cite{DePueMT}, but here we investigate an alternative route initially proposed in \cite{Sortais12}.

In this work, we demonstrate efficient loading of a single atom into a tight microtrap, sufficiently small to display collisional blockade. During collisional blockade loading, we introduce a collisional loss channel in which only one of the two colliding atoms can get lost from the trap ($2-1$ loss). This $2-1$ loss channel is a consequence of repulsive light-assisted collisions induced by blue detuned light, as also exploited in \cite{Andersen0, Carpentier13}. The idea here is that if the timescale for $2-1$ loss can be made faster than the timescale for the rapid $2-0$ loss, then it is more likely that only one of the two colliding atoms will be lost from the trap when a second atom enters. In this case, we expect that the single atom loading efficiency can exceed the $\sim 50 \%$ obtained in previous collisional blockade experiments. We achieve a loading efficiency of $\sim 80 \%$ in the collisional blockade regime. Our efficiency is limited by the rapid $2-0$ loss caused by the MOT light, which we, contrary to the method in \cite{Carpentier13}, cannot effectively eliminate since the MOT needs to be sustained during loading. Our implementation relies on using the light shifts induced by the microtrap to tune the atomic energy levels of a trapped atom to match the MOT laser frequency and hence, there are restrictions on which trap depths that can be used.
We present an analytical model that agrees with our experiment, which describes the loading dynamics in the collisional blockade regime as a function of $R$, $\gamma$ and the probabilities for both $2-0$ and $2-1$ loss. Our work highlights the qualitative difference between loading processes in the few- and many-body regimes. Contrary to the many atom regimes where increased loss rates decrease the mean atom number, we observe that the mean atom number in the microtrap grows when the two-body loss rate is increased appropriately.

The remainder of the paper is structured as follows. Section \ref{S2} reviews the concept of $2-1$ loss induced by blue detuned light-assisted collisions previously used in \cite{Andersen0, Carpentier13} and employed in this work as well. In section \ref{S3}, we introduce an analytical model for the loading dynamics in the collisional blockade regime with the presence of both $2-0$ and $2-1$ loss as well as loss not induced by collisions. Section \ref{S4} describes our experiment and section \ref{S5} presents the practical implementation and measured data, while section \ref{S6} contains discussions and conclusions.

\begin{figure}
\includegraphics[width=120mm,height=60mm]{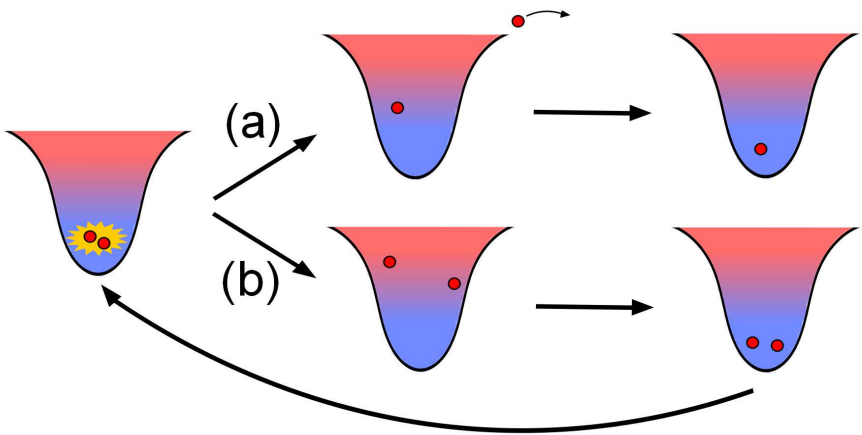}
\caption{(color online) Light-assisted collision between two atoms leading to one atom loss. Path (a): When one of the atoms has sufficient energy to escape the trapping potential, then the other atom will laser cool and stay in the trap after the collision. Path (b): If none of the atoms has sufficient energy to escape the trap, then both of them will remain in the trap with higher energy than before the collision. Laser cooling lowers the energies of the atoms and thus, the atoms are returned to the conditions prior to the collision and the process repeats.}
\label{flow}
\end{figure}

\section{Collisions between two atoms leading to one and only one atom loss} \label{S2}

A combination of laser cooling and light-assisted collisions induced by blue detuned light can lead to a two-atom collisional process where only one of the collision partners is lost ($2-1$ loss) \cite{Carpentier13}. The principle of the process is illustrated in figure \ref{flow}. 

Two atoms collide with a low (relative to the trap depth) but finite thermal energy ahead of the collision. The blue detuned light used to induce the inelastic light-assisted collision limits the maximal energy released to $h \Delta$ with $\Delta$ being the detuning of the light from the atomic resonance \cite{Thad,wiener}. Choosing $\Delta$ such that the pair of atoms after the collision has a total energy $E_p$ that is larger than the energy required for one of the two atoms to escape the optical trapping potential but insufficient for both atoms to escape, gives finite probabilities that the collision leads to none or one of the atoms escaping while the probability that both are lost vanishes \cite{Carpentier13}. If none are lost, then the laser cooling removes the energy released in the light-assisted collision, which returns the pair to the condition prior to the collision and the process can thus repeat until one of the two atoms is lost. When an atom is lost, the other atom laser cools and remains.

There are two main criteria that need to be fulfilled to achieve near unity probability for one of the two atoms to be lost while the probability that both are lost vanishes: 
\begin{enumerate}
\item \label{c1} Trap loss should arise only from light-assisted collisions induced by blue detuned light with $\Delta$ in the right range. This means that trap loss arising from light-assisted collisions induced by light sources other than the blue detuned light, and trap loss mechanisms that are not due to collisions between trapped atoms, must be eliminated. Furthermore, prior to a collision, the atoms must be in an internal ground state that has a transition such that $\Delta$ is in the right range.
    
\item \label{c2} The thermal energy of the two atoms must be large enough ahead of a collision such that the pair can have sufficient center of mass momentum to ensure that they share the released energy unevenly in order for one of the atoms to gain enough energy to be lost from the trap, while the other atom stays. At the same time, it is necessary to avoid the scenario where the pair possesses a thermal energy before a collision high enough such that both of them have enough energy to escape the trapping potential after the collision.

\end{enumerate}

The multilevel nature of atoms and the lack of accessible closed transitions without vector light shifts makes it challenging to implement the process in practice for alkali metals. Reference \cite{Carpentier13} reports an implementation for $^{85}$Rb atoms where the probability of $2-1$ loss is $93\%$, when two atoms were present in the microtrap. This was achieved as follows. The collisions were induced by a light beam (denoted the ``collision beam'') that was tuned blue relative to the $F=2$ to $F'=3$ D1 transition where both ground and excited states experience a largely $m$-independent light shift due to the trap. This ensures that the detuning of the collision beam is independent of what $m$-state the atoms are in prior to a collision, as long as they are in the $F=2$ ground state. Since the $F=2$ to $F'=3$ D$1$ transition is not closed, atoms may undergo off-resonant spontaneous Raman scattering and end up in the $F=3$ ground state. This means that without a process that returns the atoms to the $F=2$ ground state then the population would rapidly build up in the $F=3$ ground state, leading to a change of the detuning by the ground state hyperfine splitting, thereby violating criterion (i). The six cooling beams also used for the MOT in earlier stages of the experiment were continuously applied during the collision stage to optically pump the atoms back to the $F=2$ ground state by shifting the frequency of these beams near to resonance with the $F=3$ to $F'=3$ D$2$ transition for an atom at the center of the trap. Please note that the shifts of the atomic transitions due to the light forming the microtrap were substantial and therefore need to be included when tuning the laser frequencies relative to particular resonances.   

It was empirically observed that the cooling beams tuned to the vicinity of $F=3$ to $F'=3$ D$2$ transition for atoms at the center of the microtrap also provided laser cooling for atoms in the deep microtraps used. This was independent of the exact depth of the microtrap used. The cooling rate and equilibrium temperature depend on the power of the cooling beams with higher power giving larger cooling rates and lower equilibrium temperatures. The cooling rate should be high enough such that the laser cooling removes the energy released in light-assisted collisions that do not lead to trap loss, before the next collision happens. A high equilibrium temperature enhances the chance that an atom is lost in a light-assisted collision induced by the collision beam. However, once the equilibrium temperature reach around one tenth of the trap depth significant one-body loss due to the atom visiting the high energy tail of its distribution starts occurring, thereby violating criterion (i).  

While the presence of the cooling/optical pumping light is necessary, this light can also induce light-assisted collisions in addition to those induced by the collision beam. This jeopardizes the $2-1$ loss process. Contrary to the collision beam, the cooling/optical pumping light is red detuned by about $3$ GHz (the ground state hyperfine splitting) for atoms in the $F=2$ ground state. This means that light-assisted collisions induced by it generally will lead to both colliding atoms being lost from the microtrap, thus reducing the probability of $2-1$ loss. The suppression of this process can be understood by looking at the probability that two atoms in a collision event will undergo an inelastic light-assisted collision $P_I$. In the regime where this probability is low, the Landau-Zener formalism in the dressed state picture predicts that for red detuned light, it can be approximated by \cite{wiener}:
\begin{equation}
 P_I \simeq 1- \frac{P_{LZ}}{2-P_{LZ}} 
\end{equation}
where $P_{LZ}=\exp\left(\frac{-2\pi\Omega^{2}}{3v\Delta}\left(\frac{C_3}{\hbar\Delta}\right)^{1/3}\right)$ with $\Omega$ being the on resonance Rabi frequency, $\Delta$ the detuning from the free atom resonance, $v$ the the relative radial speed for the pair, and $C_3$ a constant. We see that a combination of a large detuning and low intensity of the light that induces the collisions makes $P_I$ small. Reference \cite{Carpentier13} therefore used the minimal possible intensity of the the cooling/optical pumping light required for a suitable equilibrium temperature and cooling rate. Loss due to the light-assisted collisions induced by the cooling/optical pumping light was thereby effectively suppressed, yielding a pair lifetime as long as $4$ s in the absence of blue detuned light. $2-0$ loss induced by cooling/optical pumping light was therefore not a main limitation on the probability that one and only one atom is lost.    

\section{Collisional blockade theory} \label{S3}

We seek to implement a collisional loss channel where only one of the partners are lost during loading in the collisional blockade regime. In this section we analyze the loading dynamics in the general case where loss from the microtrap can arise from three different processes. The processes are light-assisted collisions leading to one of the collision partners being lost ($2-1$ loss), light-assisted collisions leading to both atoms being lost ($2-0$ loss), and finally one-body loss ($1-0$ loss).   
Since we are considering the collisional blockade regime we will assume that the two-body loss processes ($2-0$ loss and $2-1$ loss) are much faster than the loading rate ($R$). This means that the probability for finding two or more atoms in the microtrap at a given time vanishes, and we can restrict the possible occupancies of the trap to either zero or one \footnote{We verify the validity of this assumption experimentally, as will be shown in the first paragraph of section 5.2.}. Since we are interested in regimes that yield a high probability for having one atom in the trap, we will also assume that the one body loss coefficient ($\gamma$) is smaller than or similar to $R$.  

We first consider the discrete series of events that may change the atom number in the trap. The first set of events is defined by an atom being loaded into the microtrap. If there were no atoms in the trap before the loading event then there will be one atom after it. If there was an atom present before the loading event then light-assisted collisions will cause one or both atoms to be lost almost instantaneously, leading to one or zero atoms present with probabilities $p_{2 \rightarrow 1}$ and $p_{2 \rightarrow 0}$.

In addition to loading events the atom number can change due to one-body loss when there is an atom present in the microtrap. In order to account for this process we assume that loading events as well as one-body loss events are uncorrelated and follows Poisson statistics. When there is an atom in the microtrap, then the probability that the atom number changes in an event (loading or one-body loss) is $\frac{R}{R+\gamma} p_{2 \rightarrow 0} + \frac{\gamma}{R+\gamma}$, where $\frac{R}{R+\gamma}$ is the probability that the event was a loading event and $\frac{\gamma}{R+\gamma}$ the probability that it was a one-body loss event. Similarly, one finds that the probability that the atom number does not change is $\frac{R}{R+\gamma} p_{2 \rightarrow 1}$.

When there are no atoms present in the trap the only events that can change the atom number are loading events and they will always result in that the atom number is changed to one. However, these events happen with a rate of $R$ whereas the total rate of events when there is one atom in the trap is $R+\gamma$. To simplify expressions we would like to have the same rate of events independent of whether there are zero or one atom in the microtrap. When there is zero atom in the trap, we therefore introduce a series of ``fictitious'' events with rate $\gamma$. These events do not change the atom number and therefore do not play a physical role. The probability that the atom number changes to one in an event (loading or fictitious) is therefore simply the probability that the event was a loading event $\frac{R}{R+\gamma}$. The number of atoms between events thereby form a Markov chain with transition matrix $P=\left[ \begin{array}{cc} \left(\frac{R}{R+\gamma}p_{2 \rightarrow 1} \right) & \frac{R}{R+\gamma} \\
\left( \frac{R}{R+\gamma}p_{2 \rightarrow 0}+\frac{\gamma}{R+\gamma} \right)  & \frac{\gamma}{R+\gamma} \\ \end{array} \right]$ and the probabilities for obtaining zero or one atom as a function of time is given by:
\begin{eqnarray} 
\mathbf{p}\left( t \right) &=& \left( \begin{array}{cc}
p_1\left(t \right)  \\
p_0 \left(t \right)  \\
 \end{array} \right)  \nonumber \\
 &=& \sum_{n=0}^\infty {\frac{\left(\left(R+\gamma\right) t \right)^n \exp \left(-\left(\left(R+\gamma\right) t \right) \right)}{n !} P^n \mathbf{p}\left( 0 \right)} \nonumber \\ 
 &=& \exp \left(-\left(\left(R+\gamma\right) t \right) \right) \exp \left(\left(\left(R+\gamma\right) P t \right) \right) \mathbf{p}\left( 0 \right)   \label{td}
 \end{eqnarray}
with $p_1\left(t \right)$ ($p_0\left(t \right)$) being the probability that there is one (zero) atoms in the microtrap as a function of time \footnote{One can obtain an equivalent expression to Eq. \ref{td} from a classical master equation based on the same assumptions.}. The long time steady state probabilities for detecting one (the maximal efficiency of the method) $p_1 \left(t \rightarrow \infty \right)=\frac{1}{1+\frac{\gamma}{R}+p_{2 \rightarrow 0}}$ or zero $p_0\left(t \rightarrow \infty \right)=\frac{\frac{\gamma}{R} +p_{2 \rightarrow 0}}{1+\frac{\gamma}{R} +p_{2 \rightarrow 0}}$ atoms are found as the Perron-Frobenius eigenvector of $P$ (where the expression is simplified using $p_{2 \rightarrow 0}+p_{2 \rightarrow 1}=1$). This result agrees with the previous observation that $p_1\simeq p_0 \simeq 0.5$ when $p_{2 \rightarrow 0}=1$ \cite{Schlosser01, Schlosser2} and the numerical prediction that $p_1=1$ when $p_{2 \rightarrow 0}=0$ and $R>>\gamma$ \cite{Sortais12}. We see that in order to obtain a high probability for the microtrap to be occupied by one atom, we must have $R>>\gamma$ and $p_{2 \rightarrow 0}$ must be small. Since the process is binomial, the mean atom number in the trap is simply $p_1$ and the variance is $p_1 p_0$.

\section{Experiment} \label{S4}

\begin{figure}
\includegraphics{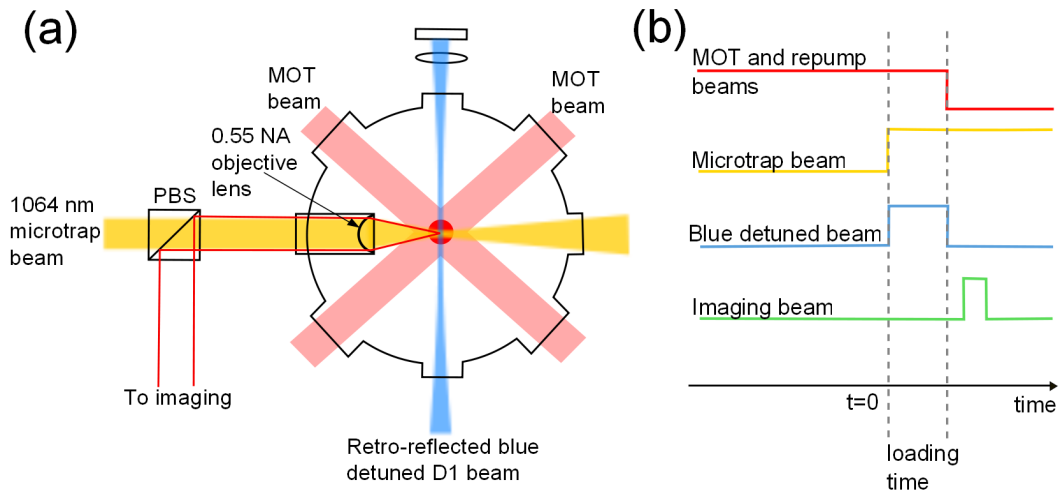}
\caption{(color online) (a) Experimental setup. The MOT and microtrap are formed inside a vacuum chamber. (b) Experimental sequence. Loading of atoms into the microtrap commences when the $1064$ nm microtrap beam is turned on.}
\label{setup}
\end{figure}

Our experimental apparatus includes a MOT for $^{85}$Rb atoms inside a vacuum chamber. The MOT beams frequency are near the cyclic $F=3$ to $F'=4$ transition of the D$2$ line and a repump beam that is on resonance with the D$2$ $F=2$ to $F'=3$ transition is included to return atoms that spontaneously decayed into the $F=2$ ground state back into the cyclic transition used for the MOT. The MOT cloud is in the focal plane of a $0.55$ numerical aperture objective lens, as shown in figure \ref{setup}(a). The microtrap is formed by the lens focusing a laser beam of $1064$ nm in wavelength to a spot size of $1$ $\mu$m inside the MOT. With the microtrap being this tight, using the loading scheme of \cite{Andersen0, Carpentier13} has lost efficiency since we are unable to load enough atoms such that we eliminate the possibility that the initial atom number is zero (see section \ref{S1}).

The $1064$nm laser beam is generated by a fiber laser. The large detuning from atomic resonances provided by this wavelength makes spontaneous scattering rates low for atoms in the microtrap. A digital control of an AOM is used for fast switching of the trap light, while the trap laser power is stabilized using feedback to an analogue modulation input. The trap laser power is set at $29.7$ mW (measured before the vacuum chamber), corresponding to a trap depth of $U_{0}=h\times47$ MHz and radial trap frequency $\omega_{r}/2\pi=115$ kHz (inferred from parametric excitation spectroscopy), unless otherwise stated.

A laser beam of variable frequency and power with $1/e^2$ radius of $150$ $\mu$m at the position of the atoms induces the desired light-assisted collisions. It is blue detuned from the $F=2$ to $F'=3$ transition on the D$1$ line and is retro-reflected to form a standing wave. The D$1$ line is chosen as the trap predominantly induces a scalar light shift on this line, leading to well defined detunings for atoms in the microtrap, which is essential for optimal performance of the method. For detection, the atoms held in the microtrap are induced to fluoresce by another D$1$ line imaging beam that is retro-reflected and mode matched to the D$1$ blue detuned beam used to induce collisions. A low light sensitive camera can detect the fluorescence collected with the objective lens. This allows us to determine the number of atoms using the method described in \cite{McGovern11}.

Figure \ref{setup}(b) illustrates the experimental sequence. Initially, the MOT is preloaded for a duration of $600~\mathrm{ms}$, unless otherwise stated. The subsequent microtrap loading rate $R$ is changed by changing the duration of this preloading stage. At time $t=0$ the microtrap and the blue detuned laser beam are turned on, and loading of atoms into the microtrap commences. After a microtrap loading stage of variable duration $t$, the loading laser beams (MOT lasers plus D$1$ blue detuned beam) are turned off to allow any untrapped atoms to escape, while the microtrap beam remains on to hold the atom in the trap. Finally, the number of atoms in the microtrap is determined. 

\section{Results} \label{S5}

\begin{figure}
\includegraphics{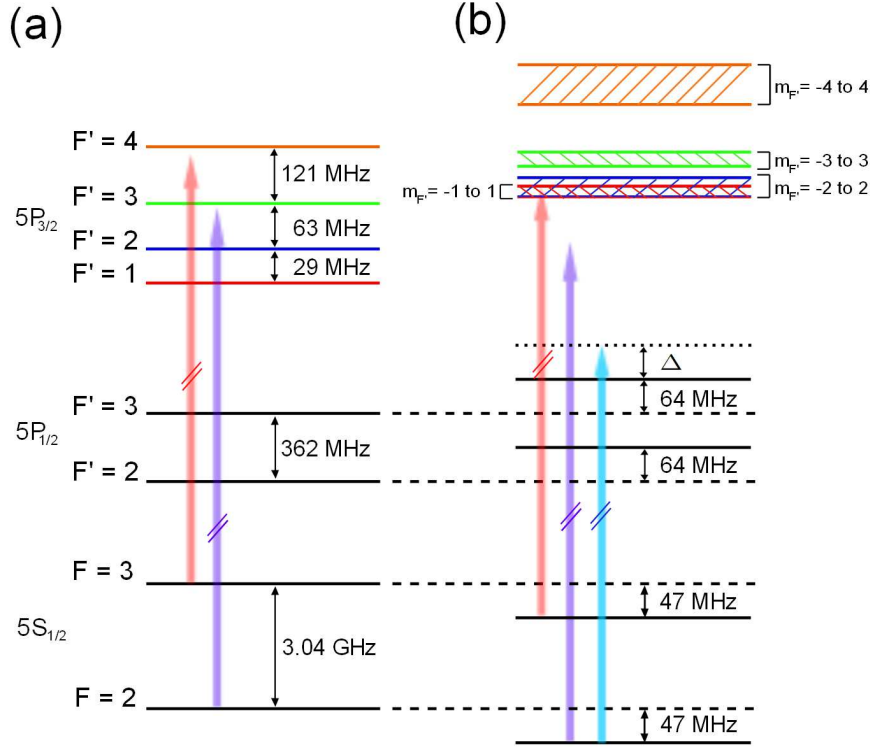}
\caption{(color online) Level diagram of $^{85}$Rb. (Not to scale) Red, purple, and blue arrows indicate the frequencies of MOT cooling light, MOT repump light, and D$1$ blue detuned beam respectively. (a) Free space level diagram. (b) Level diagram for the atom at the center of the microtrap, light shifted by the microtrap beam. The D$2$ excited states are shown as light shifted m-states manifolds.}
\label{level}
\end{figure}

\subsection{Implementation}
Our goal is to achieve a near deterministic loading of a single atom inside a tight microtrap under the collisional blockade regime. During the loading process, the frequencies and powers of the MOT beams (cooling and repump beams) that can be used are restricted since these beams need to sustain the MOT and to load atoms into the microtrap. The MOT needs to be sustained such that it provides a reservoir of atoms to load the microtrap from.
For our scheme to work efficiently, it is crucial that the rate of light-assisted collisions induced by the blue detuned laser beam (favors $2-1$ loss) exceeds the rate of those induced by the MOT light which favors $2-0$ loss. In a collision event, we therefore want the probability for light-assisted collisions induced by the red detuned beams to be low. As explained in section \ref{S2}, this probability can be determined theoretically and light-assisted collisions are unlikely for low intensity and/or a large detuning of the light that induces them \cite{wiener}.


Tuning the trap depth such that the light shift at the center of microtrap brings the $F=3$ to $F'=2$ D$2$ transition close to the cooling light frequency (red detuned from the $F=3$ to $F'=4$ D$2$ transition in free space), causes atoms at this position to be pumped into the $F=2$ ground state (see figure \ref{level}). This renders the cooling light about $3$ GHz detuned from any transitions, thereby suppressing light-assisted collisions induced by it. 
The light from the microtrap also shifts the MOT repump transition (the $F=2$ to $F'=3$ on the D$2$ line) by about $156$ MHz, consequently suppressing the ability of the MOT repump light to induce light-assisted collisions and to pump trapped atoms back to the $F=3$ ground state. 
The trap light therefore acts as a ``transparency beam" for the MOT repump light, similar to the method in \cite{Steller13}.
The microtrap itself does not induce light-assisted collisions at the timescale of our experiments due to its large detuning. To further reduce the unwanted processes we decrease the repump beam intensity to the minimal required to sustain the MOT. All these efforts, while being restricted by the fact that the MOT has to be sustained during loading, contribute towards the fulfillment of criterion (i) stated in section 2.

\subsection{Loading of one atom into a tight microtrap}

\begin{figure}
\includegraphics[width=90mm,height=65mm]{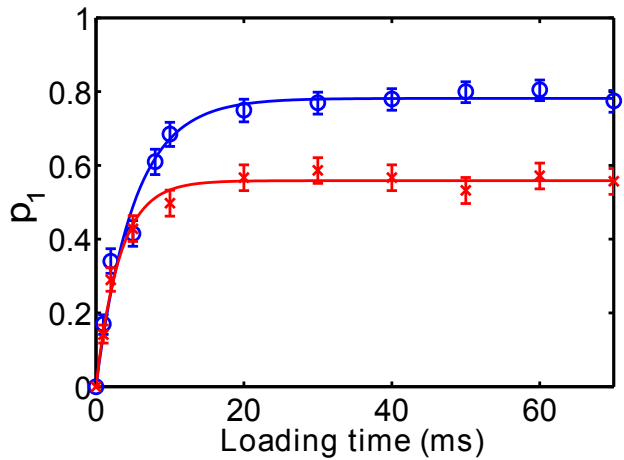}
\caption{(color online) Probability of observing one atom ($p_1$) as a function of loading time with a trap depth of $h\times47$ MHz. Probabilities are determined statistically from 200 repetitions. Blue circles are experimental data points taken with the blue detuned laser beam on. The blue solid line is a fit with Eq. \ref{td}, assuming negligible $\gamma$. The fit yields $R=156$ $s^{-1}$ and $p_{2 \rightarrow 1}=0.72$. The red crosses are data points taken with the blue detuned laser beam off. The red solid line is a fit with Eq. \ref{td}, assuming negligible $\gamma$, yielding $R=173$ $s^{-1}$ and $p_{2 \rightarrow 1}=0.21$.}
\label{fig2}
\end{figure}

Despite our efforts to suppress light-assisted collisions induced by the MOT lasers, our experiment still operates in the collisional blockade regime. This is seen from the red points in figure \ref{fig2}, which shows $p_1$ as a function of loading time without the blue detuned laser beam. The probability for observing one atom initially increases, but levels out at about $p_1\left(t \rightarrow \infty \right)=56\%$. A similar loading probability has in the past been observed in the collisional blockade regime \cite{Piotrowicz13} and is explained by a non-zero $p_{2\rightarrow 1}$ from collisions induced by MOT light \cite{Sompet13}. Additionally, several groups have employed a non-zero $p_{2\rightarrow 1}$ from collisions induced by MOT light to isolate individual atoms from small samples with loading probabilities exceeding $60\%$, see \cite{Sompet13,Roy12} and supplementary materials of \cite{Kaufman14}. In our present experiment, runs yielded either zero or one atom, confirming that the light-assisted collisions induced by the MOT lasers alone put us in the collisional blockade regime.

The loading timescale in figure \ref{fig2} is about $6$ ms. Since we observe the collisional blockade, the collisional loss timescale is much faster. The timescale for collisional loss induced by lasers other than the blue detuned laser beam was $4$ s in \cite{Carpentier13} (probably more than four order of magnitudes longer than that of the present work). This highlights one of the challenges to have the $2-1$ loss dominating over the $2-0$ loss in the collisional blockade regime, namely that the microtrap needs to be loaded and the MOT has to be sustained such that we do not have as much freedom in minimizing the $2-0$ loss induced by the MOT light. 

Despite the high rate of undesired loss induced by the MOT light, the presence of the blue detuned laser beam still enhances the single atom loading efficiency as intended. The blue points in figure \ref{fig2} show the probability of obtaining one atom when the blue detuned beam is turned on during loading (power of 12 $\mu$W, $1/e^2$ radius of 150 $\mu$m, and detuning of $68$ MHz). The solid lines are plots of Eq. \ref{td} assuming $\gamma=0$ with $R$ and $p_{2 \rightarrow 1}$ fitted. The model matches the experiment for the parameters used. Note that the $p_{2 \rightarrow 1}=0.72$ found here is expected to be lower than the $p_{2 \rightarrow 1}=0.93$ from \cite{Carpentier13} due to the afore mentioned restrictions on the MOT light parameters. As expected from Eq. \ref{td}, the MOT preload time (determines $R$) changes the timescale for the probability to reach the plateau. For the range of parameters tested, $p_1\left(t \rightarrow \infty \right)$ did not change significantly as long as the MOT preload time was above $\sim 300$ ms. This indicates that $300$ ms preload time ensures $R>>\gamma$. A simple estimate of the rate of light-assisted collisions induced by the blue detuned light gives a timescale for these to be on the order of few tens of $\mu$s, agreeing with our observation of collisional blockade for a loading timescale of $6$ ms.  

The blue detuned laser beam increases the overall two-body loss rate by introducing the $2-1$ loss channel. Given that $p_1$ is the mean atom number in the microtrap, figure \ref{fig2} demonstrates that the steady state atom number goes up when the two-body loss rate increases. This shows that the regime of $\sim 1$ atom can behave fundamentally different to loading higher atom numbers, where an increased two-body loss rate will always lead to a lower steady state atom number.

From figure \ref{fig2} we see that the plateau (the steady state) is reached at times shorter than $60$ ms. In the following we therefore investigate the probability for obtaining a single atom after $60$ ms of loading ($p_1\left(t=60~\mathrm{ ms} \right)$) as a function of different experimental parameters to gain a more detailed understanding of the loading process. The range of different parameters explored are around the parameters that give the peak probability of obtaining a single atom ($\sim80\%$).

\subsection{Trap depth}

\begin{figure}
\includegraphics[width=90mm,height=60mm]{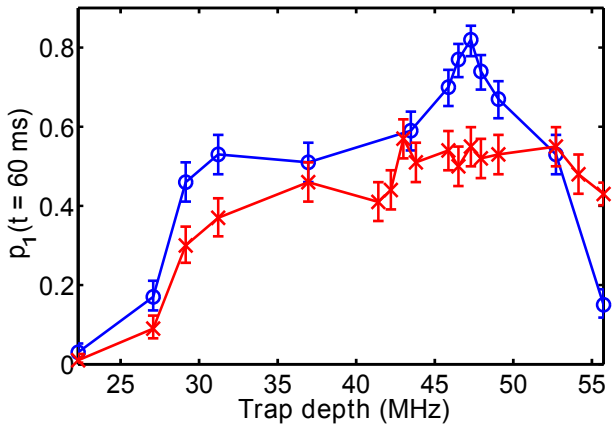}
\caption{(color online) $p_1\left( t=60~\mathrm{ms} \right)$ as a function of trap depth. Blue circles are experimental data points taken with the blue detuned laser beam. Red crosses are experimental data points taken without the blue detuned laser beam. The solid lines are guides to the eye.}
\label{fig4}
\end{figure}

Different trap depths cause different light shifts on the atomic transitions and this effect plays an important role in our experiment. We investigate how $p_1\left(t=60~\mathrm{ ms} \right)$ changes with trap depth. Without the blue detuned beam $p_1$ stays around $50 \%$ as long as the trap depth is deeper than $h\times35$ MHz for the trap depths shown. At lower trap depths, $p_1 \left( t=60~\mathrm{ms} \right)$ drops due to increasing one-body loss ($\gamma$). This happens when the equilibrium temperature of the atom is no longer negligible relative to the trap depth, leading to atom loss when they visit the high energy tail of their distribution. $p_1 \left( t=60~\mathrm{ms} \right)$ decreases at high trap depths as well (the trend continues for trap depths larger than $h \times 56~\mathrm{MHz}$). We ascribe this to the dependence of the in-trap laser cooling on trap depths. For deeper traps, the laser cooling yields a high equilibrium temperature of the atom that leads to non-negligible $\gamma$. 

With the blue detuned beam (blue circles in figure \ref{fig4}), $p_1 \left( t=60~\mathrm{ms} \right)$ increases considerably for a narrow range of trap depths around $h\times47$ MHz. At this trap depth, the trap light shifts the $F=3$ to $F'=2$ D2 transition close to resonance with the MOT cooling light (see figure \ref{level}(b)). 
This causes the atom to be depumped into $F=2$ ground state and thus reducing the unwanted light-assisted collisions induced by the red detuned MOT cooling light. 
The narrow range of trap depths over which the loading is considerably enhanced means that trap laser power drifts must be eliminated for optimal performance.
If trap depths other than those with high single atom loading probabilities are required for subsequent experiments, the microtrap laser power could be ramped adiabatically to the desired value after loading a single atom in the microtrap. 

\subsection{Detuning of the blue detuned beam}

\begin{figure}
\includegraphics[width=90mm,height=60mm]{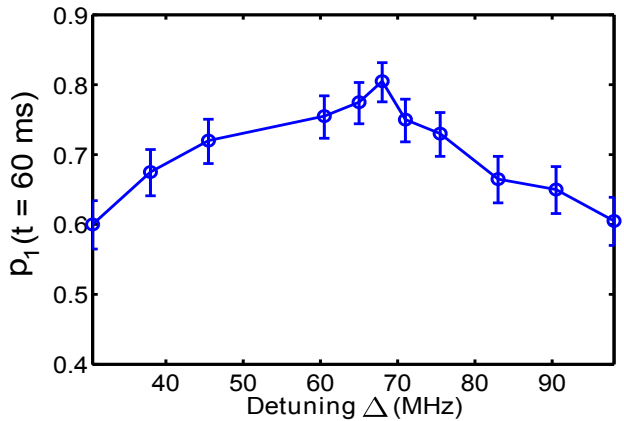}
\caption{$p_1\left( t=60~\mathrm{ms} \right)$ as a function of the blue detuned beam detuning, $\Delta$. The solid line is a guide to the eye.}
\label{fig3}
\end{figure}


The blue detuned laser beam is a crucial element in our experiment since it causes the $2-1$ loss channel needed for the enhancement of single atom loading efficiency. 
The frequency of the blue detuned laser beam determines the energy released in each inelastic collision it induces. Holding the trap depth at $h\times47$ MHz, we varied the frequency of the blue detuned beam and measure $p_1 \left( t=60~\mathrm{ms} \right)$.
Figure \ref{fig3} shows that a wide range of frequencies gives efficiencies above $50\%$ but the peak performance ($\sim 80\%$) is obtained for detunings around $68$ MHz. At this detuning, an inelastic collision releases an energy of up to $1.4$ times the trap depth, which agrees with the expectation that the optimal energy release to cause the preferred $2-1$ loss, should be enough for one of the two colliding atoms to escape but not enough for both.


From figure \ref{fig3}, we can see that the detunings $\Delta$ that give $p_1\left(t=60~\mathrm{ ms} \right)\geq0.6$ span over $60$ MHz, whereas the span of trap depths (see figure \ref{fig4}) that gives $p_1\left(t=60~\mathrm{ ms} \right)\geq0.6$ is $10$ MHz, which corresponds to a $22$ MHz light shift on the D$1$ line transitions. This indicates that the strong dependence of $p_1\left(t=60~\mathrm{ ms} \right)$ on trap depth observed in figure \ref{fig4} originates from the MOT beams strong frequency dependence on its role to perform cooling and optical pumping rather than a requirement for an accurate detuning of the blue detuned laser beam. 

\subsection{Power of the blue detuned beam}

\begin{figure}
\includegraphics[width=90mm,height=60mm]{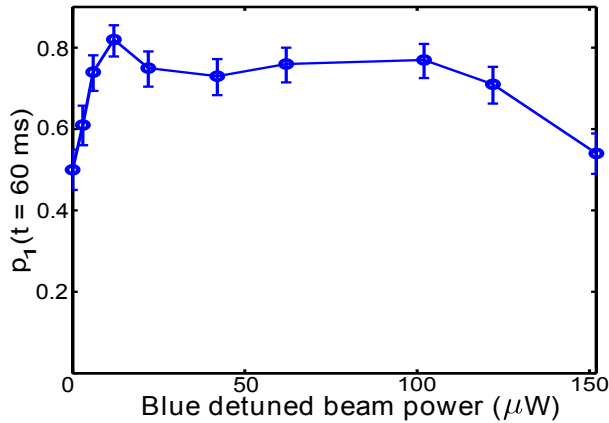}
\caption{ $p_1\left( t=60~\mathrm{ms} \right)$ as a function of the blue detuned beam power. The solid line is a guide to the eye.}
\label{fig5}
\end{figure}

The blue detuned beam power controls the rate of light-assisted collisions induced by it. Figure \ref{fig5} shows how $p_1\left(t=60~\mathrm{ ms} \right)$ changes with the power of the blue detuned beam (with detuning of $68$ MHz and trap depth of $h\times47$ MHz). Below $12$ $\mu$W we observe that $p_1\left(t=60~\mathrm{ ms} \right)$ increases with the power. At low intensities, the rate of the desired collisions induced by the blue detuned light increases with intensity. Once the blue light induced collisions have a rate significantly higher than those from other laser beams, then increasing the power no longer enhances the chance for $2-1$ loss when a second atom is loaded. We ascribe the variation of $p_1 \left(t=60~\mathrm{ ms} \right) $ for powers above $12$ $\mu$W to changes in $\gamma$ as the power of the blue detuned laser beam affects the equilibrium temperature \cite{McGovern11}.

As mentioned above we presently apply the blue detuned light in the form of a standing wave. This makes the effective power observed by the atoms drift when the phase of the standing wave drifts relative to the microtrap \cite{Carpentier13}. To check if this limits our loading efficiencies we also tried to apply the light in form of a running wave but did not observe an improvement in the loading efficiency. We therefore conclude that the drift in the phase of the standing wave is not a main limiting factor presently. 

\section{Discussion and conclusions} \label{S6}
Our experiment shows an improvement of the loading efficiency of single atoms in tight microtraps where the collisional blockade takes effect, from the previously observed $\sim50\%$ to $\sim80\%$. While the single atom loading efficiency of our work is not as high as in the related method in \cite{Carpentier13}, the present method allows for direct loading of very tight microtraps where the method of \cite{Carpentier13} is difficult to implement. Our method may therefore provide a route for futher improvements of the loading efficiency in experiments such as \cite{Xia15} that recently reported efficiencies of up to $70\%$ using similar techniques.    
The loading efficiency of the current work falls short of that of \cite{Carpentier13} primarily due to the requirement that there has to be sufficient MOT cooling and repump light to sustain the MOT during loading. One could therefore expect that the loading efficiency could be improved if the microtrap is loaded from a reservoir of atoms held in a large volume FORT instead of from a MOT.

In some applications \cite{Reitz13}, it might be necessary to load atoms into smaller traps than ours. We do not expect this to lead to lower probabilities for loading one atom. Although a smaller volume leads to a higher rate of the unwanted inelastic collisions induced by the MOT light (because the density increases), the blue light induced collision rate scales similarly. The blue detuned collisions can therefore dominate at smaller volumes as well. Our method may thus provide a route to efficient loading of a 3D optical lattice and other geometries in a similar way that arrays of microtraps were loaded in \cite{Nogrette14}. Combining this with methods for Raman sideband cooling of single atoms \cite{Kaufman12,Lukin13} may provide an exciting platform for studying coherent quantum processes in few and many-body systems.

In summary, we have demonstrated a method for efficient loading of tight optical microtraps with a single atom.
Introduction of light-assisted collisions with controlled energy release along with suppression of unwanted collision processes can significantly enhance the probability for obtaining one atom in the collisional blockade regime.
In this regime, we observe the largest of such loading probabilities reported. 
The number statistics is highly sub-poissonian with a Mandel Q parameter of about $-0.8$.
The loading dynamics is well fitted with an analytical model that describes the loading efficiency as a function of loading rate (R), single atom loss rate ($\gamma$), and $p_{2 \rightarrow 0}$ and $p_{2 \rightarrow 1}$. Our findings may be applied in quantum information processing, construction of hybrid devices, and in new ways of producing quantum degenerate gases.\\

This work was supported by the Marsden Fund Council from Government funding, administered by the Royal Society of New Zealand (Contract number UOO1320). We thank Julia Fekete for her comments on our manuscript.\\

\end{document}